\documentstyle [12pt] {article}

\begin{document}
\begin{flushright}
UDHEP-10-92\\
IUHET-228\\
\end{flushright}
\vspace*{0.5 cm}
\begin{center}
{\bf NEUTRINO MIXING DUE TO A \\
VIOLATION OF THE EQUIVALENCE PRINCIPLE} \\
\vspace*{0.5 cm}
\vspace*{0.5 cm}
{\bf J. Pantaleone}\\
\vspace*{0.5 cm}
Physics Department\\
Indiana University, Bloomington, IN 47405 \\
\vspace*{0.5 cm}
and\\
\vspace*{0.5 cm}
{\bf A. Halprin, C. N. Leung } \\
\vspace*{0.5 cm}
Department of Physics and Astronomy\\
University of Delaware, Newark, DE 19716 \\
\vspace*{0.5 cm}
{\bf ABSTRACT} \\
\end{center}

Massless neutrinos will mix if their couplings to gravity
are flavor dependent, i.e., violate the principle of equivalence.
Because the gravitational interaction grows with neutrino energy,
the solar neutrino problem and the
recent atmospheric neutrino data may be simultaneously explained by
violations at the level of \(10^{-14}\) to \(10^{-17}\) or smaller.
This possibility is severely constrained by present accelerator neutrino
experiments and will be preeminently tested in proposed long baseline
accelerator neutrino experiments.

\newpage

Several years ago, Gasperini noted that if the gravitational couplings
of neutrinos are flavor dependent, mixing will take place
when neutrinos propagate through a gravitational field \cite{G}.
Similar ideas were proposed independently by Halprin and Leung \cite{HL}.
Consequently, experiments designed to search for neutrino mixing
also probe the validity of the equivalence principle.
In this Letter, we analyze the implications of present neutrino mixing
experiments for the equivalence principle.

We consider the effects
on neutrinos when they propagate under the influence of a weak, static
gravitational field.
For simplicity, we shall assume two neutrino flavors and neglect any neutrino
masses.
Ignoring effects which involve a spin flip,
the flavor evolution of a relativistic neutrino
is quite simple (see \cite{HL,HLP} for more rigorous
derivations).  In the rest frame of a massive object, a neutrino
has the effective interaction energy
\begin{equation}
H = - 2 | \phi (r)| E (1 + f)
\label{H}
\end{equation}
where E is the neutrino energy, and
$\phi (r) = - | \phi(r) |$ is the Newtonian
gravitational potential of the object.
$f$ is a small, traceless, $2 \times 2$ matrix which parametrizes
the possibility of gravity coupling to neutrinos with
a strength different from the universal coupling,
i.e. violations of the equivalence principle.

$f$ will be diagonal in some basis which we denote as the
gravitational interaction basis (G-basis).
In that basis, $\delta \equiv f_{22} - f_{11}$,
then provides a measure of the
degree of violation of the equivalence principle.
In general, as occurs for neutrino masses, the flavor
basis or the weak interaction basis (W-basis) will not coincide with the
G-basis.  If we denote the
neutrino fields in the G-basis by $\nu_G = (\nu_1, \nu_2)$ and neutrinos in
the W-basis by $\nu_W = (\nu_e, \nu_\mu)$,  $\nu_G$ and $\nu_W$ are
related by a unitary transformation, \(U^\dagger\):
\begin{equation}
\left( \begin{array}{c}
\nu_1 \\
\nu_2 \end{array}
\right) = \left[ \begin{array}{cc}
\cos \Theta_G & - \sin \Theta_G \\
\sin \Theta_G & \ \ \cos \Theta_G \end{array}
\right] \left( \begin{array}{c}
\nu_e \\
\nu_\mu \end{array}
\right),
\label{mix}
\end{equation}
where $\Theta_G$ is the mixing angle.
Consequently when a massless neutrino propagates through a gravitational
field, flavor mixing will occur.

The idea of using degenerate particles to study possible
violations of the equivalence principle is not new.
Similar effects have been considered in
the neutral kaon system \cite{kaon} for over 30 years.
Note, however, that a violation of
the equivalence principle in the kaon system requires
that gravity couples differently to particles and antiparticles,
a violation of CPT symmetry.  This requirement is not
necessary for neutrinos. Here, gravity is coupling
slightly differently to different fermion generations.

Using Eq. (\ref{H}), we may write down the flavor evolution
equation for relativistic neutrinos propagating through
a gravitational field (with no matter present).
In the W-basis, it reads
\begin{equation}
i \frac{d}{dt} \left( \begin{array}{c}
\nu_e \\
\nu_\mu \end{array}
\right)
= E |\phi(r)| \delta \ \ U
\left[ \begin{array}{cc}
-1 & 0 \\
0 & 1 \end{array}
\right] U^\dagger \ \
\left( \begin{array}{c}
\nu_e \\
\nu_\mu \end{array}
\right).
\label{osc}
\end{equation}
where we have neglected the irrelevant term in the hamiltonian which
leads to an unobservable phase.
For constant \(\phi\), the survival probability for a $\nu_e$
after propagating a distance L is
\begin{equation}
P(\nu_e \rightarrow \nu_e) = 1 - \sin^2(2 \Theta_G) \sin^2[
 {\pi L \over \lambda } ].
\label{P}
\end{equation}
where
\begin{equation}
\lambda = 6.2 {\rm km} ( {10^{-20} \over | \phi |
\delta }) ( {10 GeV \over E})
\label{lambda}
\end{equation}
is the oscillation wavelength.

Eq. (\ref{P}) is quite similar to that for vacuum oscillations
due to neutrino masses (see e.g. \cite{KP}).
However note the linear dependence of the oscillation phase
on the neutrino energy.  For a mass,
the phase depends on \(1/E\).  Thus these
two sources of mixing can be easily distinguished by searching for
neutrino mixing at different energies.

When the neutrino propagates through matter,
then the mixing can be dramatically enhanced.
A resonance occurs when
\begin{equation}
\surd \overline{2} G_F N_e = 2 E |\phi| \delta \cos ( 2 \Theta_G )
\end{equation}
where \(G_F\) is Fermi's constant and \(N_e\) is the electron
density.
This effect is completely analogous to the
well studied situation in which the mixing
is due to neutrino masses
(\cite{MSW}, for a review, see \cite{KP}).
The survival probabilities
for gravitationally induced mixing when there is a matter
background can be obtained from those for masses
by the transformation
\begin{equation}
 {m_2^2-m_1^2 \over 4 E} \rightarrow {E | \phi| \delta} ,
\label{convert}
\end{equation}
if \(\phi\) is a constant.

The local potential, \( \phi\), enters through the
phase of the oscillations, Eq. (\ref{P}).  It
vanishes far from all sources of gravity so that results of
special relativity are recovered.
However the local value of $\phi$ is uncertain,
because the estimates tend to
increase as one looks at matter distributions at larger and larger scales.
The potential at the Earth due to the Sun is \(1\times 10^{-8}\),
that due to the Virgo cluster of galaxies is about
\(1\times 10^{-6}\) while that
due to our supercluster \cite{kaon} has recently been estimated
to be about $3 \times 10^{-5}$ (which is larger than \(\phi\)
{\it in} the Sun due to the Sun).
In what follows, we will quote values for the combined,
dimensionless parameter $|\phi| \delta$.

We now consider what the current data on neutrino mixing imply for the
parameters $|\phi| \delta$ and $\Theta_G$.
The searches for mixing can be divided into three broad categories;
laboratory experiments, atmospheric neutrino observations and
solar neutrino observations.
The latter two have shown some evidence for neutrino
mixing and will be considered first.

Solar neutrinos have been observed in four experiments \cite{solar},
their results are summarized in Table 1.
The observations are all well below the predictions of the
standard solar model \cite{BU}--an indication of mixing.
There are two mechanisms by which
neutrino mixing can give large reductions
in the solar neutrino flux,
long-wavelength oscillations or resonant conversion.
We shall consider these mechanisms separately.

If the distance between the Sun and the Earth is
half of an oscillation wavelength
and the mixing angle is large, then Eq. (\ref{P})
predicts a large reduction in the flux.
This occurs for 10 MeV neutrinos when
\(|\phi|\delta = 2\times 10^{-25}\).
Then the high energy neutrinos will be depleted but the
lower energy neutrinos will be completely unaffected.
However, the present data indicate mixing for the low energy
solar neutrinos as well, see Table 1.
A careful \(\chi^2\) analysis finds that there is no long-wavelength,
two flavor explanation of the data--it is
disfavored at the 3 standard deviation level.
This is in contrast to the normal case of mixings induced by mass
differences for which the data are well described by vacuum oscillations
of two flavors (see e.g. \cite{APP}).
The difference is due to the differing energy
dependence of the two types of mixings.
Next generation solar neutrino observations \cite{future}
will further constrain this possibility by
measuring the solar neutrino energy dependence
(SNO or Super-Kamiokande)
and by searching for seasonal variations
(BOREXINO).

Resonant conversion as the neutrinos propagate through the interior
matter of the Sun can also lead to large reductions in the flux.
Figure 1 shows the favored regions from a \(\chi^2\) fit to the
flux reduction values in Table 1
(the Kamiokande-II energy bins are included explicitly
and their overall systematic error is correctly accounted for \cite{solar}).
An analytical expression was used to describe the
resonant conversion survival probability
\cite{KP,HLP}, based on only the Sun's gravitational potential.
The average gravitational potential in the sun,
$|\overline{\phi}| = 4 \times 10^{-6}$, was used to normalize Fig. (1).
If a constant potential from larger scales dominates over this,
then the allowed regions in Figure 1 are slightly
elongated about the mean $|\overline {\phi}| \delta$
by approximately a factor of 3.
Next generation solar neutrino observations
will test these regions \cite{HLP}
by measuring the solar neutrino energy dependence
(SNO or Super-Kamiokande),
by looking for day-night variations,
and by performing a neutral current measurement of the
solar flux (SNO).

Two experiments \cite{atmos} recently found that there is a
relative depletion of \(\nu_\mu\) to \(\nu_e\)
in the flux of low energy atmospheric neutrinos.
This depletion may be the result of flavor mixings.
The energy of the atmospheric neutrinos is typically about $0.5$ GeV.  The
propagation length now varies from 20 to 10,000 kilometers.  This
corresponds to values of $| \phi | \delta$ from about $6 \times 10^{-20}$
to about $10^{-22}$.  Although the data can be explained by either
$\nu_\mu-\nu_e$ mixing or $\nu_\mu-\nu_\tau$
mixing, and so do not necessarily probe the same
parameters as solar neutrinos, it is encouraging that a
range of \(|\phi|\delta\) can account for the solar neutrino data and
the atmospheric neutrino data simultaneously \cite{HLP}.  This may signal
a possible breakdown of the principle of equivalence.

Numerous laboratory experiments have been performed
searching for neutrino mixing.  They have not found definitive evidence for
mixing, so these experiments eliminate ranges of the
gravitationally induced mixing parameters.
The most stringent laboratory limits come from
"appearance" experiments with the largest values of \(E \times L\).
These occur in experiments using
the beams of neutrinos produced by accelerators \cite{limits},
where the neutrino energies
are tens of GeV and propagation lengths are as long as a kilometer.
Because most mixing experiments only analyze the data
in terms of \({L \over E}\), the relevant quantity for neutrino masses,
the results are not exactly transferable to gravitationally
induced mixing.
{}From the published descriptions of the experiments,
we have {\it estimated} the average value of \(E \times L\)
to derive the limits shown in the top part of Fig. (2).
Our estimates are probably accurate up to factors of 3 in
\( | \phi | \delta \).

Most of the large mixing angle region which
solves the solar neutrino problem is
eliminated by the current laboratory bounds.
But the lower part of this solution, and the small
mixing angle region, are still allowed by the present accelerator data.
Only relatively small improvements in the current bounds are needed
to constrain these regions.

New accelerator neutrino experiments, with
baselines of hundreds or thousands of kilometers,
are under active consideration at the present time.
In the lower part of Fig. (1) are shown estimates of the accessible
parameter region achievable by two of the proposed
experiments \cite{lbane}, FNAL to Soudan 2 and FNAL to DUMAND.
At these long distances matter effects
are becoming important \cite{ELAN}, as is apparent
in the difference between \(\nu_\mu\) and \( \overline{\nu}_\mu\) regions.
The energy distribution of the neutrino event
rate is taken from calculations for
a short baseline experiment using the FNAL main injector.
Also, similar neutrino energies, intensities and
propagation lengths are available to planned
next generation atmospheric neutrino detectors
(such as DUMAND, AMANDA, etc.). They may be able to probe parameter regions
similar to those shown for the accelerator experiments.
Thus there are many planned and proposed experiments which can extend
the tests of the equivalence principle to values of \(|\phi| \delta\)
far below the present accelerator limits.

In conclusion, the degeneracy of neutrinos enables
tests of the equivalence principle which are far more sensitive than
those using "normal" matter \cite{normal}.
The present solar neutrino data suggest a
violation of the equivalence principle at the level of
\( 2 \times 10^{-19} < |\phi|\delta <  3 \times 10^{-22}\).
The atmospheric neutrino data also suggest a possible breakdown
of the equivalence principle at this same level.
This possibility can be independently checked by long-baseline
accelerator neutrino experiments, which can reach down to
\(|\phi|\delta \approx 10^{-24}\). The violation of the
equivalence principle is introduced on purely phenomenological grounds.
Such a violation would indicate a breakdown in general relativity
or the existence of additional long range tensor interactions.

We thank M. Butler, S. Nozawa, R. Malaney and A. Boothroyd for communicating
preliminary results from their analysis.
CNL wishes to thank Fermilab for their hospitality
and the University of Delaware
Research Foundation for partial support.
JP thanks Brookhaven and the Queen's University Summer Institute
for their hospitality.
This work was supported in part by the U.S. Department of Energy under
Grants No. DE-FG02-84ER40163 and DE-FG02-91ER40661.

\raggedbottom

\raggedbottom
\newpage

\setlength{\baselineskip}{18pt}
%
%

Table 1.  Results of the solar neutrino experiments \cite{solar}.
The flux is given as a fraction of the standard solar model \cite{BU}
prediction.
\\

\begin{tabular}{| l | l | l | l |} \hline
Experiment & Process & E\(_{threshold}\) & Expt./SSM \\ \hline
Davis et al. & \(\nu_e + ^{37}\)Cl\( \rightarrow e + ^{37}\)Ar
   & 0.81 MeV  & 0.27 \(\pm\) 0.04 \\
Kamiokande-II & \(\nu + e \rightarrow \nu + e \)
   & 7.5 MeV  & 0.46 \(\pm\) 0.05 \(\pm\) 0.06 \\
SAGE & \(\nu_e + ^{71}\)Ga\( \rightarrow e + ^{71}\)Ge
   & 0.24 MeV  & 0.44 \(^{+0.13}_{-0.18}\) \(\pm\) 0.11 \\
GALLEX & \(\nu_e + ^{71}\)Ga\( \rightarrow e + ^{71}\)Ge
   & 0.24 MeV  & 0.63 \(\pm\) 0.14 \(\pm\) 0.06 \\ \hline
\end{tabular}
\raggedbottom


\newpage

\begin{center}
{\bf FIGURE CAPTION} \\
\end{center}
\vspace*{0.6cm}

\noindent {\bf Fig. 1.}  $\chi^2$ plot showing regions of $|\overline
          {\phi}| \delta$ versus \(\sin^2 2\Theta_G\) allowed by the solar
          neutrino data in Table 1 at 90\% (solid lines) and 99\%
          (dotted lines) confidence level, assuming two flavors
          and \(\delta > 0\). \\

\noindent {\bf Fig. 2.}  Upper contours: Regions of $|\phi| \delta$
	  versus \( \sin^2 2\Theta_G \)
          that are excluded by accelerator neutrino data \cite{limits}.
          Lower contours: Regions probed by proposed \cite{lbane}
	  long-baseline experiments, assuming sensitivity to
	  10\% $\nu_\mu$ ($\overline{\nu_\mu}$) disappearance.
	  The outer dashed (inner dash dot) curve is for a
	  \(\nu_\mu\) (\(\overline{\nu}_\mu\)) beam from
	  FNAL \(\rightarrow\) DUMAND.
	  The inner dashed (dot) curve is for a
          \(\nu_\mu\) (\(\overline{\nu}_\mu\)) beam from
          FNAL \(\rightarrow\) Soudan 2.

\end{document}